**Detangling Extrinsic and Intrinsic Hysteresis for Detecting Dynamic Switch of Electric Dipoles using Graphene Field-Effect Transistors on Ferroelectric Gates**

*Chunrui Ma[‡]*[a], Youpin Gong[‡a], Rongtao Lu[a], Emery Brown[b], Beihai Ma[c], Jun Li[b], and Judy Wu*[a]*

[a] Department of Physics and Astronomy, University of Kansas, Lawrence, Kansas, 66045, USA

[b] Department of Chemistry, Kansas State University, Manhattan, Kansas 66506, USA

[c] Energy Systems Division, Argonne National Laboratory, Argonne, Illinois 60439, USA

[‡]These authors contributed equally to this work.

*Address correspondence to: **chunrui.ma@gmail.com; jwu@ku.edu**

**Abstract:** A transition in source-drain current vs. back gate voltage ($I_D-V_{BG}$) characteristics from extrinsic polar molecule dominant hysteresis to anti-hysteresis induced by an oxygen deficient surface layer that is intrinsic to the ferroelectric thin films has been observed on graphene field-effect transistors on $Pb_{0.92}La_{0.08}Zr_{0.52}Ti_{0.48}O_3$ gates (GFET/PLZT-Gate) during a vacuum annealing process developed to systematically remove the polar molecules adsorbed on the GFET channel surface. This allows detangle of the extrinsic and intrinsic hysteresis on GFET/PLZT-gate devices and detection of the dynamic switch of electric dipoles using GFETs, taking advantage of their high gating efficiency on ferroelectric gate. A model of the charge trapping and pinning mechanism is proposed to successfully explain the transition. In response to pulsed $V_{BG}$ trains of positive, negative, as well as alternating polarities, respectively, the source-drain current $I_D$ variation is instantaneous with the response amplitude following the $I_D-V_{BG}$ loops measured by DC $V_{BG}$ with consideration of the remnant polarization after a given $V_{BG}$ pulse when the gate electric field exceeds the coercive field of the PLZT. A detection sensitivity of around 212 dipole/$\mu m^2$ has been demonstrated at room temperature, suggesting the GFET/ferroelectric-gate devices provide a promising high-sensitivity scheme for uncooled detection of electrical dipole dynamic switch.

## 1. Introduction

Graphene has remarkable electronic properties including high carrier mobility, broad-band optical transmittance, chemical inertness and mechanical flexibility, making it a promising candidate for electronic, photonic and optoelectronic applications.[1-6] In particular, the single atomic layer structure of graphene implies graphene is an excellent sensor material with extremely high sensitivity. In field-effect device geometry, graphene illustrates high bipolar susceptibility to the gate electric field, which provides a unique sensing scheme for biosensing,[7, 8] chemical and gas sensing,[9] and photo-detection through ionic liquid gate mediated light-to-gate field coupling.[10] Among several characteristic directly associated to the sensor performance, dynamic response is an important parameter that defines quantitatively the speed and profile a sensor responds to the signal. Understanding the intrinsic as well as extrinsic mechanisms that affect the dynamic response of the graphene-based devices is hence essential to their applications in high speed devices such as memory devices employing a scheme of graphene field-effect transistors on ferroelectric gate (GFET/FE-gate).[11-13] In this type of devices, the "0" and "1" states defined by values of the source-drain current $I_D$ in the GFETs, are manipulated by controlling the electric dipole switch in the FE-gate based on the intrinsic ferroelectric hysteresis. The dynamic response of the $I_D$ therefore determines the operation speed as well as the yield of the memory devices. However, the dynamic response to the gate field in the GFET/FE-gate devices has not been studied systematically, primarily due to the difficulties associated to the co-existence of the anti-hysteresis induced by oxygen deficient surface layer of ferroelectric gates (For simplicity, "intrinsic ferroelectric hysteresis" will be used in the rest of the text) and extrinsic one caused by charged polar molecules adsorbed on graphene.[14] The ferroelectric hysteresis stems from the dipole alignment in nonlinear response to an applied electric field, resulting from a strong dependence of the polarization on the magnitude, initial value, and sweeping direction of the electric field. Since it is the polarization at the graphene/gate interface that affects the $I_D$ in a GFET/FE-gate device via either n-type or p-type charge doping of graphene, the ferroelectric hysteresis is manifested in a shift of the Dirac point (at which a minimum $I_D$ is obtained) when the gate voltage $V_{BG}$ is swept up (towards more positive $V_{BG}$ direction, or forward) or down (towards more negative $V_{BG}$ direction or backward). While the ferroelectric hysteresis is intrinsic, Dirac point shift to both directions, either as expected[11, 13, 15] or opposite (anti-hysteresis)[12, 16, 17], have been reported on GFETs/FE-gate devices. Further complication arises from the interference of the extrinsic hysteresis attributed to air molecules and contaminants possibly attached on graphene during GFETs fabrication, as manifested in $I_D$ hysteresis in GFETs as $V_{BG}$ is swept in forward and backward directions on non-ferroelectric $SiO_2$ gates[14] not to mention the sample-to-sample inconsistency. Understanding the intrinsic ferroelectric dynamic response in a GFET/FE-gate device therefore requires detangling the intrinsic and extrinsic hysteresis effects. In this paper, we present a systematic study of the dynamic response of GFETs on ferroelectric $Pb_{0.92}La_{0.08}Zr_{0.52}Ti_{0.48}O_3$ (PLZT) thin film gate (GFET/PLZT-gate) to $V_{BG}$ pulse trains by first detangling the intrinsic and extrinsic hysteresis. A vacuum annealing process was developed to remove the adsorbed polar molecules. A transition from a co-existence of intrinsic and extrinsic hysteresis states to the anti-hysteresis state induced by an oxygen deficient surface layer that is intrinsic to the ferroelectric thin films has been demonstrated with increasing the annealing period. Dynamic $I_D$ in response to electric dipole switch driven by positive, negative, as well as alternating polarity $V_{BG}$ pulse trains, respectively, was studied after the extrinsic hysteresis was removed. The resemblance between the pulsed and DC $I_D-V_{BG}$ characteristic illustrates that a quantitative correlation between $I_D$ and the gate dipole alignment can be established in a dynamic fashion below and above the coercive field of the ferroelectric PLZT gate.

PLZT was selected considering its advantages as a gate dielectric for room temperature applications. Compared with un-doped PZT, Lanthanum doping can effectively reduce oxygen vacancy, decrease leakage current, and lower the fatigue and domain pinning.[18,19,20] Especially, PLZT is in the ferroelectric tetragonal region with remarkable piezoelectric effect, high dielectric permittivity around 1300, high breakdown electric field around 2.0 MV/cm, as well as the slim hysteresis loop beneficial for low-loss dynamic energy storage applications.[21,22] It also has a suitable Curie temperature of about 180 °C,[23] preventing phase change from ferroelectric to dielectric under the thermal budget of 150 °C typically applied for fabrication of GFET/PLZT-gate devices.[24] In a recent work, we have shown the domain mobility and nonlinearity are further enhanced in epitaxial PLZT thin films.[25] This together with the reduced coercive field makes epitaxial PLZT films an ideal gate candidate for GFETs with high gating efficiency, large-scale tunability of polarization, and dynamic switch of the electric dipoles using low gate voltages.[25]

## 2. Experimental

### 2.1 Fabrication of epitaxial PLZT film

Epitaxial (001)-oriented PLZT films are fabricated using pulsed laser deposition (PLD) at the similar condition as in an earlier report.[25] Briefly, the PLD of epitaxial PLZT films of 500 nm in thickness were carried out at 650 $^{o}$C under 225 mTorr oxygen partial pressure on conductive (001) Nb:SrTiO$_3$ substrate with a KrF excimer laser (wavelength of 248 nm and pulse width of 25 ns). The average laser pulse energy density was 2 J/cm$^2$ and repetition rate was 5 Hz. After the deposition, the PLZT films were *in-situ* annealed at 600 Torr oxygen partial pressure for 30 minutes to reduce oxygen vacancies in the films before cooling down to room temperature. X-ray diffraction (Bruker AXS D8 diffraction system) was employed to confirm the epitaxial growth of the PLZT films on Nb:SrTiO$_3$.[22] A Radiant Technologies Precision Premier II tester was applied to analyze the ferroelectric properties of PLZT thin film capacitors.

### 2.2 Graphene fabrication and transfer

Graphene were grown at ~1000 $^{o}$C in a chemical vapor deposition (CVD) system on commercial polycrystalline copper foils of 25 μm in thickness. Graphene sheets of typically 1x1 cm$^2$ in dimension were transferred onto the PLZT films using a modified procedure based on what we reported earlier.[26-28] In order to transfer the graphene films onto a PLZT thin film, poly-methyl methacrylate (PMMA) was spin-coated on one side of the graphene sample before it was submerged in copper etchant (CE100) for removal of the copper foil. After this procedure was completed, the graphene samples were rinsed with deionized (DI) water for multiple times before being transferred onto PLZT thin films. This process is critical to eliminating surface contaminants on graphene which contribute partially to the extrinsic hysteresis in $I_D-V_{BG}$ loops of GFET/PLZT-gate devices. After cleaning, the graphene sheets were suspended in DI water for transfer to solid surfaces. In particular, to transfer a thoroughly cleaned graphene sheet onto a PLZT film, the PLZT film was immerged into the DI water to engage one side of the suspended graphene. Lifting the graphene/PLZT film assembly out of the water was carried out carefully in order to smoothly engage the entire graphene sheet onto the PLZT surface. This step was found critical to thoroughly remove residues that can be trapped at the interface between graphene and PLZT and hence induce extrinsic effect to the GFET characteristics. After the transfer, the samples were baked in air at 150 °C for one hour to eliminate moisture, which was followed by immersing them in acetone in order to remove the PMMA on the graphene. Isopropyl Alcohol rinse was employed afterwards to remove residues on the surface of the graphene.

**2.3 GFETs device fabrication and electrical transport measurement**

Source and drain electrodes were defined in the first photolithography, which was followed with electron-beam evaporation of 2 nm titanium/88 nm of gold and liftoff. In the second photolithography, GFETs were defined and fabricated with reactive ion etch (RIE) in oxygen plasma at 20 W RF power. The oxygen partial pressure was 6.7 mTorr and the RIE time was 150 seconds. The GFETs devices investigated in this work have the same channel width and length of 20 μm. Special steps were employed in our graphene transfer process to avoid contact of graphene samples to any surfaces to minimize contamination of the graphene surface (see details in Experiment). In addition, the samples were kept in high vacuum better than $5\times10^{-6}$ Torr for extended period of more than 24 hours in our probe station before electrical transport measurements, which was found effective in eliminating contamination on graphene surface.

## 3. Results and discussion

Figure 1a-d include an optical image and Raman maps of D, G and 2D peaks (WiTec Alpha 300 confocal Micro-Raman system), respectively, on the graphene channel in a representative GFET/PLZT-gate device. Negligible D peak was detected as illustrated in Figure 1b and 1e primarily at the edges of the GFET channel, suggesting the CVD graphene of high quality remained intact during graphene transfer and GFET fabrication process. The intensity of the 2D peak is around twice of that of the G peak, which is anticipated on the single-layer graphene. The inset of Figure 1e depicts a GFET/PLZT-gate device schematically. The dielectric constant and loss of the epitaxial PLZT thin films were characterized in capacitor form as function of electric voltage across the device, as shown in Figure S1. The charging-discharging loops taken on a 500 nm thick PLZT film are shown in Figure S2.

In order to detangle the effects of the intrinsic ferroelectric hysteresis and the extrinsic hysteresis by external molecules adsorbed to graphene, a series of $I_D-V_{BG}$ loops at different $V_{BG, max}$ values were measured on the GFET/PLZT-gate devices while the environment was systematically varied from in air to high vacuum for variable periods during the vacuum annealing at room temperature. Figure 2 compares four groups of the $I_D-V_{BG}$ loops measured on the same GFET/PLZT-gate device in air (Figure 2a), after 10 minutes pumping to low vacuum of $2.8 \times 10^{-5}$ Torr (Figure 2b), and after 1 day (Figure 2c) and 3 day (Figure 2d) pumping, respectively, in high vacuum of $<4 \times 10^{-6}$ Torr. It is evident that the hysteresis behavior changed dramatically as the environment was varied. Most strikingly, the hysteresis direction reversed, as exemplified from the switch of the respective positions of the Dirac points "F" (at $V_{Dirac,F}$) and "B" (at $V_{Dirac,B}$), respectively, on the forward and backward branches of the $I_D$-$V_{BG}$ loops. When measured in air as shown in Figure 3a, the $V_{Dirac,F}$ locates at right-hand side of the $V_{Dirac,B}$, indicating the former is at a more p-doped state as compared to the latter. This behavior sustained as the $V_{BG, max}$ was varied in the range of 0.5 to 3.0 V covering both below and above the coercive voltage $V_c \sim 0.6$V (defined from coercive field $E_c$ of about 12 kV/cm, see details in Figure 4d of the PLZT gate. In contrast, the relative positions of the $V_{Dirac,F}$ and $V_{Dirac,B}$ switched to the opposite positions with $V_{Dirac,F}$ representing a more n-type doping state after the sample was in high vacuum for 3 days as shown in Figure 2d. An additional fundamental difference between these two cases is the occurrence of the hysteresis only when $V_{BG, max}$ is above $V_c$ in Figure 2d, which is anticipated from the appearance of the ferroelectric remnant polarizations when the applied electric field is above the coercive field. This result suggests the dominance of the extrinsic hysteresis mechanism before an extended vacuum "cleaning" was applied. This argument is supported by a transition between the two cases observed with increasing vacuum annealing time as illustrated in the hysteresis trend evolution from Figure 2a, through Figure 2b and 2c, and finally to Figure 2d. The fact that the extrinsic hysteresis could be reduced and ultimately removed by vacuum annealing implies that the molecules of contaminants adsorbed on graphene are most probably responsible for the extrinsic hysteresis in GFETs and also explains an unexpected hysteresis observed on GFET/SiO$_2$–gate devices in air[14]. In this work, we have found that the vacuum annealing is effective to remove the extrinsic hysteresis and the removal is permanently by one-time extended pumping of 3 days in high vacuum. No extrinsic hysteresis was observed even if the GFET/PLZT-gate devices were measured again in air after the vacuum annealing, suggesting the adsorbed polar molecules responsible for the extrinsic hysteresis are most probably those from chemicals involved in graphene transfer and GFET-PLZT-gate device fabrication, rather than air molecules. While this extrinsic polar molecule effect may be eliminated through improvement of the graphene transfer and GFET fabrication processes, the result in this work has illustrated it can be effectively removed using a robust vacuum post-annealing process.

To quantify the hysteresis of both the intrinsic and extrinsic origins, Figure 3a displays the Dirac point positions as function of the $V_{BG,max}$ values in the four cases shown in Figure 2: in air (squares), after 10 minutes of pumping to low vacuum of 2.8x10$^{-5}$ Torr (triangles), and after 1 day (circles) and 3 days (diamonds), respectively, in high vacuum of <4x10$^{-6}$ Torr. When the extrinsic mechanism dominates (Figure 3a, first column from left), neither $V_{Dirac,B}$ nor $V_{Dirac,F}$ have a clear trend as the $V_{BG,max}$ range is varied. In contrast, the absolute values of both the $V_{Dirac,B}$ and $V_{Dirac,F}$ increase monotonically with $V_{BG,max}$ after the extrinsic mechanism is removed (Figure 3a, last column from left), which is anticipated from the increase of the ferroelectric remnant polarization with increasing $V_{BG,max}$ at above $V_c$. When $V_{BG,max}$ was below $V_c$, the same $V_{Dirac,B}$ and $V_{Dirac,F}$ were observed as $V_{BG,max}$ because electric dipole switch in a ferroelectric material occurs only at $E>E_c$ (or $V_{BG,max} > V_c$), resulting in no remnant polarization and therefore no ferroelectric hysteresis *at* $V_{BG,max} < V_c$. This certainly is not the case when extrinsic hysteresis dominates where large hysteresis is observable in the entire range of the $V_{BG,max}$, both below and above $V_c$. Figure 3b further compares several $I_D$-$V_{BG}$ curves taken at $V_{BG,max}$=0.1, 0.2 and 0.4 V (much below $V_c$), respectively, on the same GFET/PLZT-gate sample before and after the complete vacuum annealing. Hysteresis persists in the former while disappears in the latter at $V_{BG,max}$ below $E_c$ as expected.

Figure 3c depicts the evolution of the $\Delta V_{Dirac}$ vs. $V_{BG,max}$ curve shapes during the transition from the extrinsic hysteresis dominant (in air, squares), to mixed (10 minutes to 1 day in vacuum, circles and triangles), and to intrinsic ferroelectric hysteresis dominant (3 days in high vacuum, diamonds) state. A qualitative change can be clearly seen in the $\Delta V_{Dirac}$ vs. $V_{BG,max}$

curves from an inverted bell shape on the negative $\Delta V_{Dirac}$ side (below the dashed line) to monotonically increasing on the positive $\Delta V_{Dirac}$ side (above the dashed line). In particular, a systematic upward shift of the $\Delta V_{Dirac}$ vs. $V_{BG,\,max}$ curve from negative to positive is clearly seen as the adsorbed species are gradually detached, rendering dominance of the intrinsic ferroelectric hysteresis effect ultimately. The magnitude of the shift is large and much greater than the ultimate $\Delta V_{Dirac}$ value (diamonds). This implies that the dynamic response in a GFET to the gate field could be highly ambiguous without detangling the intrinsic and extrinsic hysteresis mechanisms. Eliminating the extrinsic hysteresis mechanisms is hence essential to detection of the dynamic response of the GFET/FE-gate devices.

Although intrinsic, the ferroelectric PLZT gate induced an anti-hysteresis in the $I_D-V_{BG}$ characteristic of the GFET/PLZT-gate devices, which has been reported previously on GFETs on other oxide ferroelectric gates.[12, 16, 17] The anti-hysteresis may be attributed to an surface oxygen-deficient layer occurs on most oxide ferroelectric thin films.[29] In the absence of this interfacial layer, the polarization $P$ at the graphene/FE-gate interface will be determined by the $P-E$ curve measured on parallel plate capacitors (Figure 4d) at a given maximum applied voltage $V_{BG,\,max}$. As depicted schematically in Figure. S2c-g, the forward (backward) sweep of the $V_{BG} > V_c$ generates an upward (downward) oriented $P$ with positive charge (negative charge) at the graphene/FE-gate interface that makes GFET more n-doped (p-doped). The presence of the "intrinsic" oxygen-deficient surface layer of oxide ferroelectric films introduces an interface between graphene and FE-gate.[30] The oxygen vacancies in this interfacial layer behave like donor ions and such mobile charges could cause a transition from a p-type inside the film to n-type at the ferroelectric film surface.[31] In the ferroelectric thin film capacitors with metal/FE interfaces (such as graphene/PLZT) being included, a compensating double layer of space charges may form at the interface and its contribution has been confirmed by extrapolating the thickness of the FE layer to zero in the reciprocal capacitance vs FE thickness curve.[29] Consequently, anti-hysteresis appears in $I_D-V_{BG}$ characteristics with ferroelectric oxide gates as depicted schematically in Figure 3d. This argument seems consistent with the majority experimental observations of anti-hysteresis in $I_D-V_{BG}$ characteristic of the GFET/FE-gate devices, whether using exfoliated graphene (assuming cleaner graphene/FE interface) or CVD graphene (with possible chemical residues at the graphene/FE interface during graphene transfer) with ferroelectric oxide gates[12, 16, 17] as well as that of hysteresis in $I_D-V_{BG}$ characteristic if more insulating polymer ferroelectric PVDF gate.[11, 13] Interestingly, this oxygen-deficiency interfacial effect was found sensitive to device operation temperature and a transition from anti-hysteretic $I_D-V_{BG}$ characteristic at high temperatures to hysteretic one was observed at 100 K for GFET/(Ba,Sr)TiO$_3$-gate[16] and at 20 K for GFET/multilayer PbTiO$_3$/SrTiO$_3$-gate devices.[30] While protecting FE gate surface may have beneficial effects, eliminating the oxygen-deficient surface layer that is "intrinsic" to most ferroelectric oxides requires a systematic work.[29]

When polar molecules are adsorbed to graphene, the adsorbed polar molecules respond to the ferroelectric polarization controlled by the $V_{BG}$ and contribute to the effective polarization sensed by the GFET as shown schematically in Figure 3e. At an initial state $V_{BG}=0$ before a gate voltage is applied, the polar molecules on both the inner and outer surface of graphene may cause a switch from n-doped to a p-doped graphene at a high enough molecule density. Consequently, the Dirac point appears at $V_{BG}>0$ on the forward $I_D$-$V_{BG}$ branch. At $V_{BG,max}>V_c$ where the sweeping direction changes from forward to backward sweeping, the hole traps pin the polar molecules on the outer surface of graphene (facing air) and the molecule dipoles with negative charge toward graphene make the graphene becomes more n-type. As a result, the Dirac point is shifted towards the more negative direction when sweep backward, which explains the "hysteretic" switching, observed in Figure 2a. Such a contribution is quantitatively reduced during vacuum annealing when the density of adsorbed polar molecules on graphene is reduced, and ultimately minimized, causing a switch from polar molecules "hysteresis" to ferroelectric "anti-hystereis" of the $I_D$-$V_{BG}$ characteristic as observed in Figure 2a-d. In addition to the relative Dirac points position switch on forward and backward $I_D-V_{BG}$ branches, the polar molecules also affect the $I_D-V_{BG}$ characteristics quantitatively. As shown in Figure 2a, the pinning of the polar molecules on top of the GFET after the $V_{BG}$ sweeping direction reversal leads to a plateau of $I_D$ of almost constant value in a large range of $V_{BG}$. The width of the $I_D$ plateau reduces systematically with vacuum annealing period (Figure 2b-c) and become negligible in Figure 2d, leading to the anticipated monotonically increasing bi-polar $I_D$-$V_{BG}$ curves in absence of polar molecules.

The dynamic response of the GFET was investigated after a thorough vacuum cleaning by applying various pulsed square-waves of $V_{BG}$ were applied to the PLZT gate in the GFET/PLZT devices. Figure 4a-c include three different $V_{BG}$ pulse trains (blue) with pulse period in the range of 84 ms−1.0 s applied to the PLZT gate: positive pulses with increasing amplitude in Figure 4a, negative pulses with increasing amplitude in Figure 4b, and pulses with alternating polarities and increasing amplitude in Figure 4c. For a positive (negative) $V_{BG}$ pulse in Figure 4a (Figure 4b), the polarization will only take a few specific discrete points of A-to-C-to-D (A-to-E-to-D), instead of a continuous trajectory on the P-E loops defined in Figure 4d. The amplitude of the $V_{BG}$ pulse, or the $V_{BG, max}$ at Point C (or Point E), determines the remnant polarization at Point D (or Point A). On the other hand, a pair of pulses of alternating polarities and the same amplitude would take discrete points of A-C-D-E-A on the P-E loop in Figure 4d. The measured $I_D$ values (black) in response to the $V_{BG}$ pulses (blue) is also pulsed and synchronized well with the $V_{BG}$ pulses as shown in Figure 4a-c, suggesting the $I_D$ responses with a considerable fast speed to the $V_{BG}$ modulation, which is anticipated since dipole switch can typically happen in sub-ps time frame. In the case of the positive $V_{BG}$ pulses, the $I_D$ pulse amplitude increases monotonically with the $V_{BG}$ pulse in the range of 0-2.0 V as shown in Figure 4a, implying an increased charge (holes) doping in graphene as a result of increased upward polarization across the PLZT gate. The $I_D$ responses to negative $V_{BG}$ pulses show a more complicated pattern since the Dirac point of ~-0.6 V locates in the range of $V_{BG}$ at 0 to −2.0V. A quantitative discussion will be given in Figure 5. The case of a pair of positive and negative pulses of the same amplitude applied to the PLZT gate may be viewed as a combination of the positive and negative pulse trains shown in Figure 4a and b.

Quantitatively, the increase in $I_D$ pulse amplitude is highly nonlinear in response to the linear increase of the $V_{BG}$ pulse amplitude. This is not surprising since PLZT, or ferroelectric materials in general, has strong nonlinear dielectric constant peaking at approximately zero field and decreasing monotonically with increaing applied electric field strength.[25] This agrees well with the reduced $I_D$ increase rate with increasing $V_{BG}$ pulse amplitude. The sensitivity of GFET to the PLZT polarization may be estimated from the detectable $I_D$ at the smallest $V_{BG}$ pulse amplitude, which increases monotonically with $V_{SD}$. At $V_{SD}$ = 100 mV, it can detect $I_D$ at the $V_{BG}$ ~1 mV. The surface polarization ($P_{GFET}$) on the PLZT-gate is estimated ~0.0034 μC/cm$^2$ from the $I_D$−$V_{BG}$ characteristic on the GFET/PLZT-gate device, or an equivalent dipole density change of 212/μm$^2$ using the following Equation 1:

$$\Delta Q = \frac{L \times \Delta I_D}{\Delta V_{SD} \times W \times \mu} \quad (1)$$

where $L$ and $W$ are the length and width of the graphene channel, respectively, and $\mu$=87.5 cm$^2$/Vs is the average charge mobility of graphene on PLZT. The $P_{GFET}$ value is substantially smaller than the polarization of $P_{Cap}$=0.014 µC/cm$^2$ calculated from the *P-E* loops on PLZT capacitors shown in Figure 4d, which is not surprising considering the screening effect of the interfacial layer on a PLZT gate. Interestingly, the ratio $P_{GFET}/P_{Cap}$~24% in this work is consistent to the polarization ratio measured in PbZr$_{0.2}$Ti$_{0.8}$O$_3$ capacitors with and without graphene being inserted between the Pd top electrode and PbZr$_{0.2}$Ti$_{0.8}$O$_3$.[15] Since PbZr$_{0.2}$Ti$_{0.8}$O$_3$ is an oxide ferroelectric material similar to the PLZT used in this work, it is most probably that the interfacial screening layer on PbZr$_{0.2}$Ti$_{0.8}$O$_3$, as well as on PLZT, is affected by graphene while the microscopic mechanism requires further investigations.

Figure 5a-c plot the $I_D$ value at $V_{BG}$ =0 after each consecutive pulse in the $V_{BG}$ pulse trains of positive, negative and alternating polarities shown in Figure 4a-c, respectively. Obviously, the $I_D$ at $V_{BG}$ =0 maintains a constant value when the amplitude of the pulse is below the $V_c$ due to a negligible *P-E* hysteresis under $V_{BG} < V_c$ as shown in Figure 4d. In contrast, the $I_D$ value at $V_{BG}$ =0 changes systematically once the pulse amplitude exceeds $V_c$ as a consequence of the remnant polarizations. The increasing positive remnant polarization with increasing $V_{BG}$ pulse amplitude shifts the $I_D$ value at $V_{BG}$ =0 towards more p-doped due to the anti-hysteresis as shown in Figure 5a. Similarly in Figure 5b, a more n-doped shift in the $I_D$ value at $V_{BG}$ =0 was observed in the case negative $V_{BG}$ pulse train at $|V_{BG}| > V_c$. In addition, a train of pairs of positive and negative pulses of the same amplitude with increasing amplitude shown in Figure 4c can be explained by combining the result for a positive and a negative $V_{BG}$ pulses. The measured $I_D$ values at the peaks of the positive (solid red) and negative (open red) $V_{BG}$ pulses are compared in Figure 5d with their counterpart of DC $V_{BG}$ (black) with both the forward and backward branches included at the same $V_{BG,max}$ of 2.0 V. In comparison of the positive pulsed $I_D$–$V_{BG}$ curve with the forward DC branch, the former takes a steeper increase than the latter while both are following a qualitatively similar trend of field-induced hole-doping. The comparison between the negatively pulsed $I_D$–$V_{BG}$ curve with the backward DC branch shows a similar trend with the same minimum $I_D$. Interestingly, the $I_D$–$V_{BG}$ curve (blue) taken with paired opposite pulses follows well the combined $I_D$–$V_{BG}$ curve of the positive (solid red) and negative (open red) pulse trains.

## 4. Conclusions

In conclusion, a vacuum annealing process has been developed to systematically remove the polar molecules adsorbed on the channel surface of the GFET/PLZT-gate devices. A transition in $I_D$–$V_{BG}$ characteristics from hysteretic to anti-hysteresis has been observed. The former could be attributed to the charge trapping and pinning mechanism of the polar molecules in response to the $V_{BG}$ reversal from forward to backward direction or vice versa, while the latter is anticipated from the intrinsic bulk ferroelectric hysteresis in combination of an oxygen-deficient surface layer. Detangling the extrinsic and intrinsic hysteresis mechanisms not only shed lights on the ambiguous $I_D$–$V_{BG}$ characteristics observed on GFETs/ferroelectric-gate devices, but also provides feasibility in detection of dynamic electric dipole switch in the ferroelectric PLZT gate using GFETs. Both DC and pulsed $V_{BG}$ were applied to probe the $I_D$ response to the electrical dipole alignment and switch in the steady-state and dynamic-state modes respectively. An excellent agreement has been obtained between the steady-state and dynamic $I_D$−$V_{BG}$ characteristics and the nonlinear behavior of $I_D$ is quantitatively correlated to that of the electric dipoles in the different ranges of electric field below and above the coercive field of ferroelectric PLZT. With demonstration of high sensitivity of 212 dipole/μm$^2$ to a dipole switch, GFETs are promising to detecting the electric dipole dynamics and correlations at room temperature.

**Acknowledgments**

The authors acknowledge support in part by NASA contract No. NNX13AD42A, ARO contract No.W911NF-12-1-0412, and NSF contracts Nos. NSF-DMR-1105986 and NSF-EPSCoR-0903806, and matching support from the State of Kansas through Kansas Technology Enterprise Corporation.

**Supporting Information Available:** Description of the dielectric constant and loss of the epitaxial 500nm PLZT thin films; The current density as function of applied electric voltage measured on a PLZT thin film in parallel-plate capacitor geometry; Schematic diagrams for the electrical dipole alignment and switch in the different ranges of the applied electric field.

**Notes and references**

1. P. Avouris, Z. Chen and V. Perebeinos, *Nature Nanotechnology*, 2007, **2**, 605-615.
2. A. K. Geim and K. S. Novoselov, *Nature Materials*, 2007, **6**, 183-191.
3. F. Miao, S. Wijeratne, Y. Zhang, U. C. Coskun, W. Bao and C. N. Lau, *Science*, 2007, **317**, 1530-1533.
4. C. Girit, V. Bouchiat, O. Naamanth, Y. Zhang, M. F. Crommie, A. Zetti and I. Siddiqi, *Nano Letters*, 2009, **9**, 198-199.
5. C. Neumann, C. Volk, S. Engels and C. Stampfer, *Nanotechnology*, 2013, **24**, 444001.
6. R. Lu, C. Christianson, A. Kirkeminde, S. Ren and J. Wu, *Nano Letters*, 2012, **12**, 6244-6249.
7. Y. Y. Shao, J. Wang, H. Wu, J. Liu, I. A. Aksay and Y. H. Lin, *Electroanal*, 2010, **22**, 1027-1036.

8.  N. Mohanty and V. Berry, *Nano Letters*, 2008, **8**, 4469-4476.
9.  F. Schedin, A. K. Geim, S. V. Morozov, E. W. Hill, P. Blake, M. I. Katsnelson and K. S. Novoselov, *Nature Materials*, 2007, **6**, 652-655.
10. Guowei Xu, Rongtao Lu, Jianwei Liu, Hsin-Ying Chiu and J. Wu, *Advanced Optical Materials*, 2014, **2**, 729.
11. Y. Zheng, G. X. Ni, C. T. Toh, M. G. Zeng, S. T. Chen, K. Yao and B. Ozyilmaz, *Appl Phys Lett*, 2009, **94,** 163505.
12. E. B. Song, B. Lian, S. M. Kim, S. Lee, T. K. Chung, M. S. Wang, C. F. Zeng, G. Y. Xu, K. Wong, Y. Zhou, H. I. Rasool, D. H. Seo, H. J. Chung, J. Heo, S. Seo and K. L. Wang, *Applied Physics Letters*, 2011, **99,** 042109.
13. S. Raghavan, I. Stolichnov, N. Setter, J. S. Heron, M. Tosun and A. Kis, *Appl Phys Lett*, 2012, **100**, 023507.
14. H. M. Wang, Y. H. Wu, C. X. Cong, J. Z. Shang and T. Yu, *Acs Nano*, 2010, **4**, 7221-7228.
15. C. Baeumer, S. P. Rogers, R. J. Xu, L. W. Martin and M. Shim, *Nano Letters*, 2013, **13**, 1693-1698.
16. A. Rajapitamahuni, J. Hoffman, C. H. Ahn and X. Hong, *Nano Letters*, 2013, **13**, 4374-4379.
17. X. Hong, J. Hoffman, A. Posadas, K. Zou, C. H. Ahn and J. Zhu, *Appl Phys Lett*, 2010, **97**, 033114.
18. S. Y. Liu, L. Chua, K. C. Tan and S. E. Valavan, *Thin Solid Films*, 2010, **518**, E152-E155.
19. M. Tyunina, J. Levoska, A. Sternberg and S. Leppavuori, *J Appl Phys*, 1998, **84**, 6800-6810.
20. B. Yang, T. K. Song, S. Aggarwal and R. Ramesh, *Applied Physics Letters*, 1997, **71**, 3578-3580.
21. S. Tong, B. H. Ma, M. Narayanan, S. S. Liu, R. Koritala, U. Balachandran and D. L. Shi, *ACS Appl. Mater. Interfaces*, 2013, **5**, 1474-1480.
22. E. Brown, C. Ma, J. Acharya, B. Ma, J. Wu and J. Li, *ACS Appl. Mater. Interfaces*, 2014, **6**, 22417.
23. M. Narayanan, B. H. Ma, U. Balachandran and W. Li, *J Appl Phys*, 2010, **107,** 024103.
24. J. W. Suk, A. Kitt, C. W. Magnuson, Y. F. Hao, S. Ahmed, J. H. An, A. K. Swan, B. B. Goldberg and R. S. Ruoff, *Acs Nano*, 2011, **5**, 6916-6924.
25. C. Ma, B. Ma, S.-B. Mi, M. Liu and J. Wu, *Applied Physics Letters*, 2014, **104**, 162902.
26. J. Liu, R. Lu, G. Xu, J. Wu, P. Thapa and D. Moore, *Advanced Functional Materials*, 2013, **23**, 4941-4948.
27. B. A. Ruzicka, S. Wang, J. Liu, K.-P. Loh, J. Z. Wu and H. Zhao, *Optical Materials Express*, 2012, **2**, 708-716.

28. G. Xu, J. Liu, Q. Wang, R. Hui, Z. Chen, V. A. Maroni and J. Wu, *Advanced Materials*, 2012, **24**, OP71-OP76.
29. M. Dawber, K. M. Rabe and J. F. Scott, *Rev Mod Phys*, 2005, **77**, 1083-1130.
30. M. H. Yusuf, B. Nielsen, M. Dawber and X. Du, *Nano Lett*, 2014, **14**, 5437-5444.
31. J. Nowotny and M. Rekas, *Ceram Int*, 1994, **20**, 251-255.

**Figure 1**. a) Optical image of a GFET/PLZT-gate device with dashed lines showing the edges of graphene channel of 20 μm (length) x 20 μm (width) between the source (S) and drain (D) electrodes. b-d) Raman maps of D, G and 2D peaks, respectively, on the same GFET channel depicted in Figure 1a. e) A Raman spectrum of the graphene transferred on the PLZT thin film gate. The inset shows GFET/PLZT-gate device schematic. The excitation laser is at 488 nm wavelength.

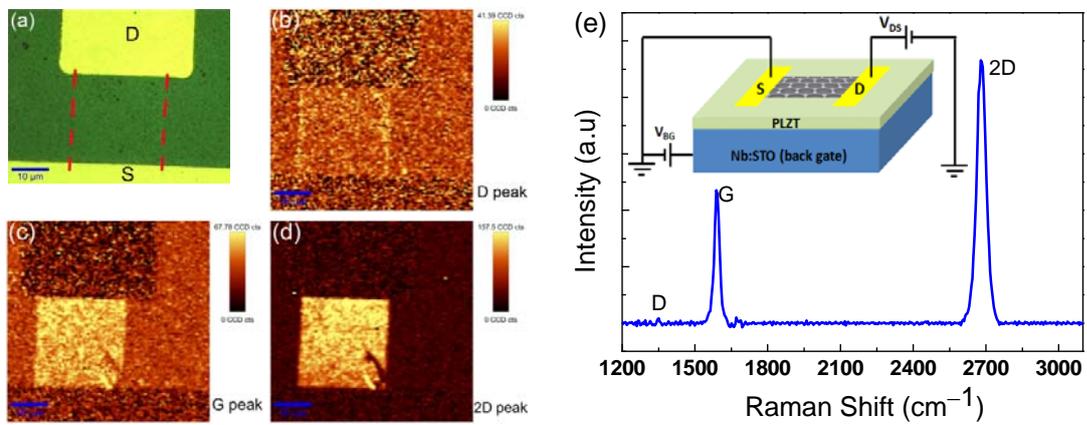

**Figure 2.** Sweeping-forward and sweeping-backward $I_D$-$V_{BG}$ curves at variable $V_{BG,\ max}$ under different measurement conditions: a) in air, b) in vacuum pumping for 10 min, c) keeping in vacuum for 1 day and d) keeping in vacuum for 3 days. The Dirac points on the sweeping-forward and sweeping-backward branches are labeled with "F" and "B", respectively.

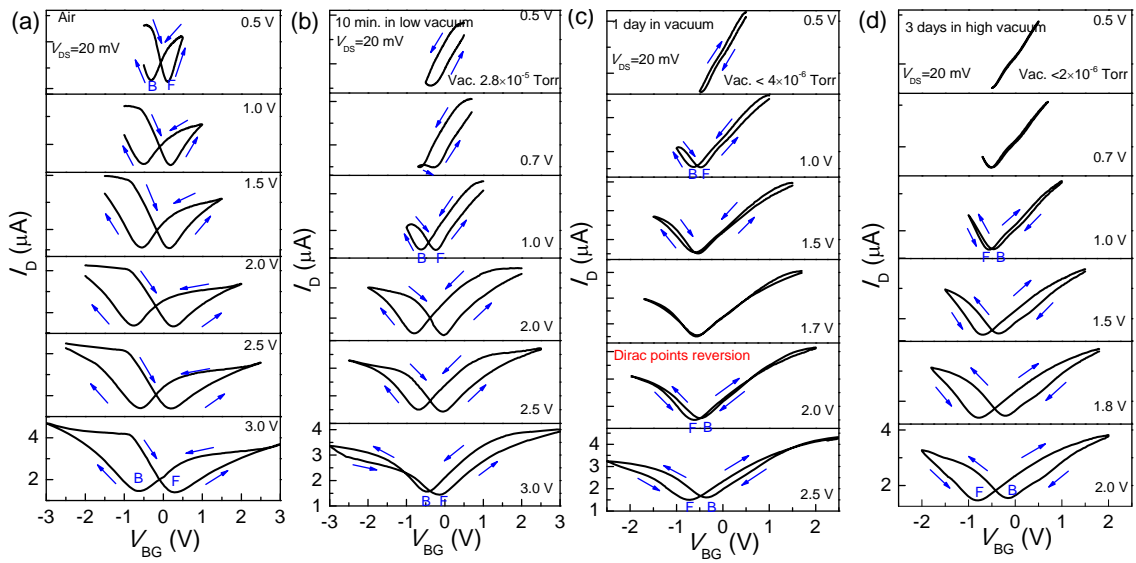

**Figure 3.** a) $V_{Dirac,B}$ (red) and $V_{Dirac,F}$ (black); and b) $\Delta V_{Dirac}$ as function of the $V_{BG,max}$ taken in the four conditions shown in Figure 2. c) Sweeping-forward and sweeping-backward $I_D$-$V_{BG}$ curves at $V_{BG,max}$ =0.1, 0.2, 0.4 mV (below $V_c$) under two experimental conditions of Figure 2b and 2d respectively. d-e) Schematic descriptions of the polarization near graphene during a forward sweep of $V_{BG}$ followed by a backward one (sequence along the direction of the black arrow) in the ferroelectric gate with oxygen deficient interface with GFET without d) and with e) the presence of the polar molecules adsorbed on graphene.

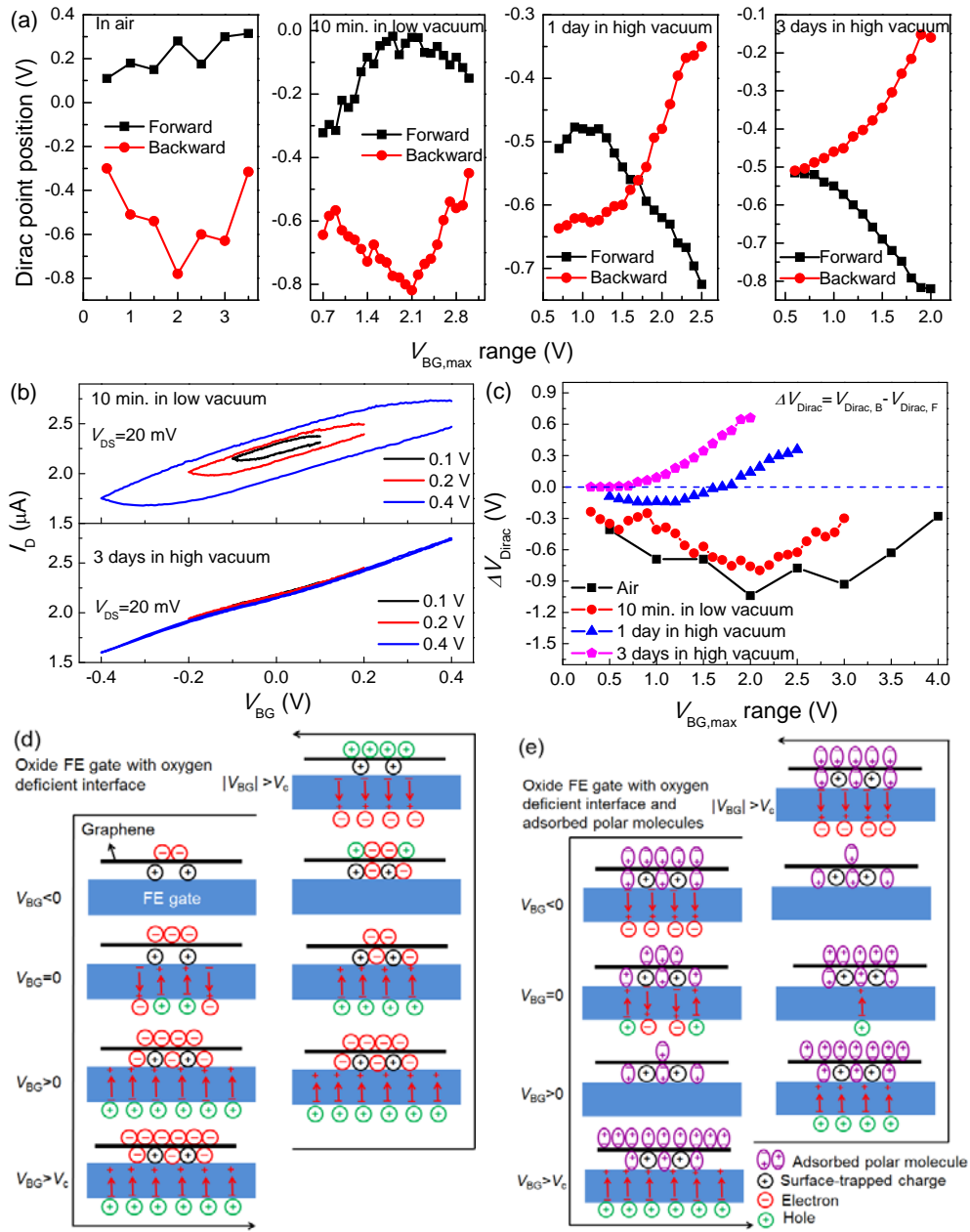

**Figure 4.** The $I_D$ in response to a) positive pulses of $V_{BG}$; b) negative pulses of $V_{BG}$; and c) $V_{BG}$ pulse pairs of alternating polarities with increasing pulse amplitude from 0 to 2.0 V at 200 ms period and 50% duty cycle. d) P-E hysteresis loops measured in 500 nm thick PLZT capacitors with the maximum applied voltages of ±0.5 V (black), ±1.0 V (blue) and ±4.0 V (red), respectively.

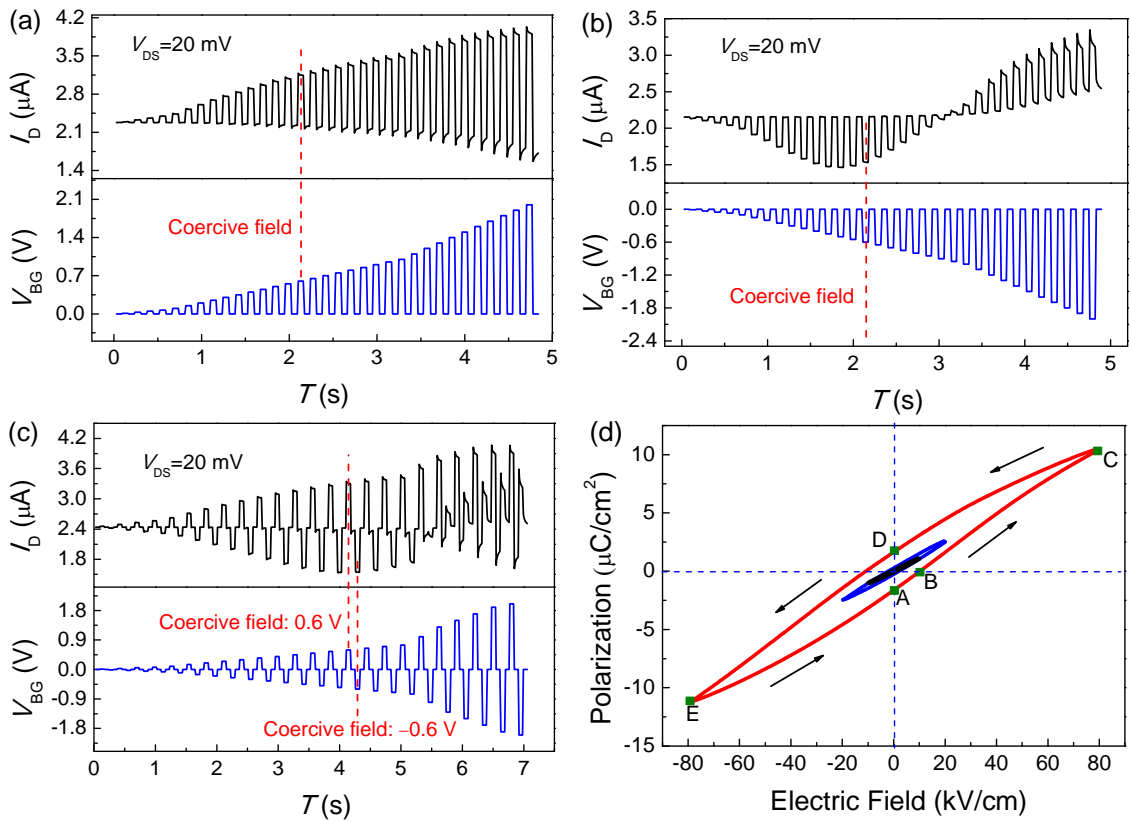

**Figure 5.** The $I_D$ at $V_{BG}$=0 when $V_{BG}$ is back to 0 V after each pulse for positive pulses a) in Figure 4a, b) negative pulses in Figure 4b and c) alternating pulses in Figure 4c, correspondingly. d) Comparison of $I_D$ curves measured at DC and pulsed $V_{BG}$ in the range of ±2.0 V, $V_{SD}$=20 mV for all curves.

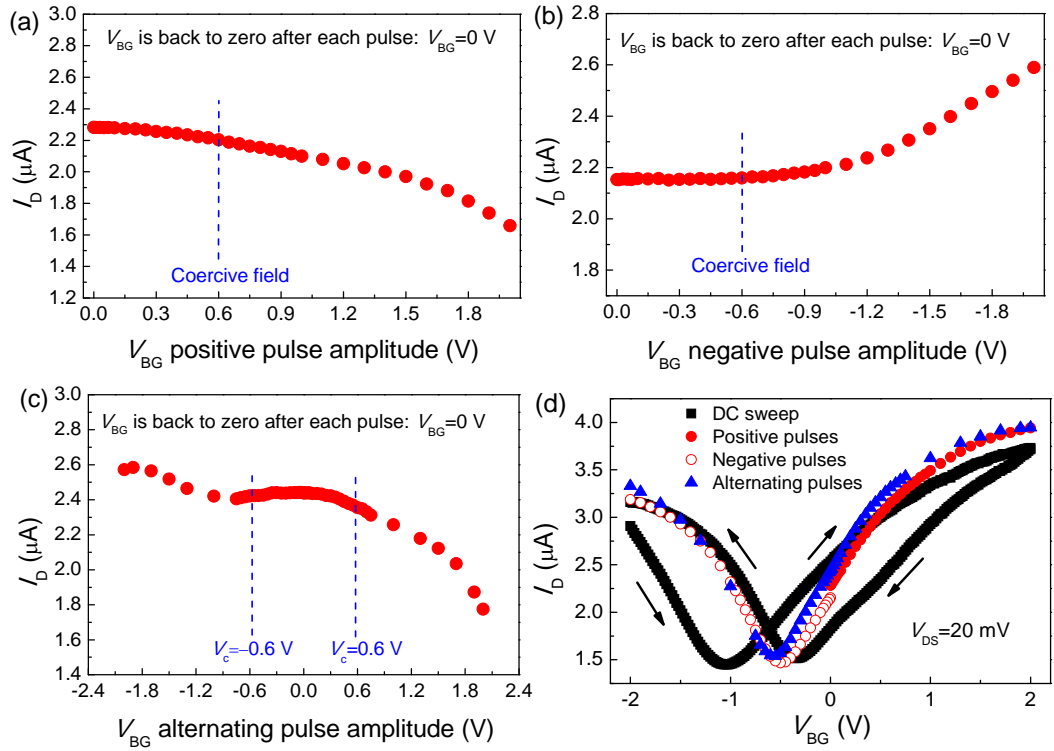

**TOC**: Graphene field-effect transistors are used to detect ferroelectric electric dipole switch driven by $V_{BG}$ pulses with positive, negative, as well as alternating polarities, respectively.

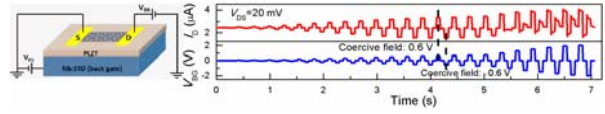